\documentclass[10pt,twocolumn,letterpaper]{article}

\usepackage{3dv}
\usepackage{times}
\usepackage{epsfig}
\usepackage{graphicx}
\usepackage{amsmath}
\usepackage{amssymb}
\usepackage{booktabs} % for much better looking tables
\usepackage{array} % for better arrays (eg matrices) in maths
\usepackage{paralist} % very flexible & customisable lists (eg. enumerate/itemize, etc.)
\usepackage{authblk}
\usepackage{verbatim} % adds environment for commenting out blocks of text & for better verbatim
\usepackage{subfig} % make it possible to include more than one captioned figure/table in a single
\usepackage{diagbox}
\usepackage{multirow}
\newcommand{\norm}[1]{\left\lVert#1\right\rVert}

\usepackage[pagebackref=true,breaklinks=true,letterpaper=true,colorlinks,bookmarks=false]{hyperref}

\threedvfinalcopy % *** Uncomment this line for the final submission

\def\threedvPaperID{155} % *** Enter the 3DV Paper ID here

\ifthreedvfinal\fi
\begin{document}
%%%%%%%%% TITLE
\title{Simultaneous Hand Pose and Skeleton Bone-Lengths Estimation from a Single Depth Image}
%\author{Jameel Malik\\
%Institution1\\
%Institution1 address\\
%{\tt\small firstauthor@i1.org}
%% For a paper whose authors are all at the same institution,
%% omit the following lines up until the closing ``}''.
%% Additional authors and addresses can be added with ``\and'',
%% just like the second author.
%% To save space, use either the email address or home page, not both
%\and
%Ahmed Elhayek\\
%Institution2\\
%First line of institution2 address\\
%{\tt\small secondauthor@i2.org}
%}

%\author{Jameel Malik\\\small{TU, Kaiserslautern}\\\small{NUST, Pakistan}\\\tt\small{jameel.malik@dfki.de} \and Ahmed Elhayek\\\small{TU, Kaiserslautern}\\\tt\small{Ahmed.Elhayek@dfki.de} \and Didier Stricker\\\small{TU, Kaiserslautern}\\\tt\small{Didier.Stricker@dfki.de}}
\author[1,2]{Jameel Malik}
\author[1]{Ahmed Elhayek}
\author[1]{Didier Stricker}
%\vspace{-20mm}
\affil[1]{\it Department Augmented Vision, DFKI Kaiserslautern, Germany}
%\affil[2]{\it TU Kaiserslautern, Germany}
\affil[2]{\it NUST-SEECS, Pakistan}
\affil[ ]{\normalsize\tt\textit {\{jameel.malik,ahmed.elhayek,didier.stricker\}@dfki.de}}

\clearpage\maketitle
\thispagestyle{empty}

%\thispagestyle{empty}
% This setup helps to train CNN more efficiently on datasets captured with different depth cameras and contain many subjects. 
%%%%%%%%% ABSTRACT
\begin{abstract}
Articulated hand pose estimation is a challenging task for human-computer interaction. The state-of-the-art hand pose estimation algorithms work only with one or a few subjects for which they have been calibrated or trained. Particularly, the hybrid methods based on learning followed by model fitting or model based deep learning do not explicitly consider varying hand shapes and sizes. In this work, we introduce a novel hybrid algorithm for estimating the 3D hand pose as well as bone-lengths of the hand skeleton at the same time, from a single depth image. The proposed CNN architecture learns hand pose parameters and scale parameters associated with the bone-lengths simultaneously. Subsequently, a new hybrid forward kinematics layer employs both parameters to estimate 3D joint positions of the hand. For end-to-end training, we combine three public datasets NYU, ICVL and MSRA-2015 in one unified format to achieve large variation in hand shapes and sizes. Among hybrid methods, our method shows improved accuracy over the state-of-the-art on the combined dataset and the ICVL dataset that contain multiple subjects. Also, our algorithm is demonstrated to work well with unseen images. 
\end{abstract}
% We also demonstrate the effectiveness of our approach on depth images captured from unseen subjects. 
% Among hybrid methods, our proposed algorithm achieves state-of-the-art performance on public datasets
% However, publically available real hand pose datasets captured from depth cameras lack in big variation in hand shapes and sizes i.e. subjects, complexity hand poses and number of depth frames. Therefore,
%%%%%%%%% BODY TEXT
\vspace{-7mm}
\section{Introduction}

Human hand is an example of complex articulable object that exhibit many degrees of freedom (DoFs), self similarities, self occlusions and constrained parameters. With the arrival of commodity depth cameras and notable progress in machine learning in the past few years, the research on human hand tracking and pose inference has gained more popularity and has become an active area of research. 
% Since, a large number of possible hand poses exist, the task of modeling human hand offers a real challenge specially for real-time and low latency applications. % keskin2012hand,tang2013real,xu2013efficient,

Mainly, three approaches exist for hand pose estimation. First, generative (model based), second, discriminative (learning based) and third, hybrid approach. Generative method starts by defining a calibrated hand model geometry and optimize an energy function to obtain the hand pose parameters \cite{oikonomidis2011efficient,qian2014realtime,tagliasacchi2015robust,tang2015opening}. These methods achieve higher accuracy at the cost of complex energy functions optimizations \cite{sharp2015accurate}. On the other hand, discriminative approach tries to infer a coarse hand pose based on already learned information from single depth, RGB-D or RGB images during training \cite{oberweger2015hands,sinha2016deephand,zimmermann2017learning,guo2016two,deng2017hand3d,zhang2016learning,li20153d,taylor2016efficient}. Recently published CNN-based methods such as hierarchical tree-like structured CNN \cite{madadi2017end}, multiview-CNN \cite{ge2016robust,ge2017robust} and region ensemble network \cite{guo2017region}  have shown significant improvement in accuracy over their counterpart, random forest based methods \cite{sun2015cascaded,wan2016hand, xu2017lie}. Despite of the fact that direct joints regression using CNN has achieved higher accuracy over other existing methods and our approach, the estimated pose is coarse and do not exploit hand geometry i.e. kinematics and physical constraints. Hence, independent learning of hand joints is most likely to produce invalid hand poses especially during tracking. In hybrid method, the pose inference obtained from discriminative method can be fed as coarse input to a generative process to get refined hand pose \cite{tompson2014real,oberweger2015training,sridhar2013interactive,sridhar2016real,ye2016spatial}. Particularly, Zhou et al. \cite{zhou2016model} propose an efficient model based deep learning approach as an alternative to generative post-processing step in hybrid methods. However, a big limitation of this work is an assumption of a fixed bone-lengths hand model geometry during end-to-end training. Clearly, this limitation restricts the generalization of this approach over different hand shapes and sizes. Our idea is to estimate not only the 3D hand pose but also the bone-lengths of hand skeleton at the same time. To the best of our knowledge, this problem has never been explicitly addressed before. So, we introduce a novel hybrid algorithm which simultaneously estimates the 3D hand pose and bone-lengths of hand skeleton. To this end, hand scale parameters are learned to facilitate the end-to-end training process of model based deep learning approach thereby, leading to promising results for 3D hand pose estimation. 

In order to show the validity of our approach, a hand pose dataset with large variation in hand shapes and sizes is necessary. Several real hand pose datasets are publicly available, but individually, these datasets lack in varying hand shapes and sizes of subjects, number of original depth images and complexity of hand poses \cite{barsoum2016articulated}. Therefore,  we combine most commonly used real hand pose datasets and convert them into a single unified format, we call {\it HandSet}. We summarize our main contributions as follows:
\begin{enumerate}
  \item A novel hybrid approach for simultaneous estimation of 3D hand pose and bone-lengths of hand skeleton.
  \item A combined real hand pose dataset that offers large variation in hand shapes and sizes, increased number of pre-processed depth frames from different depth cameras and complex hand poses. The dataset will be publicly available. 
\end{enumerate}

%\begin{figure}
%\begin{center}
%%\fbox{\rule{0pt}{2in} \rule{.9\linewidth}{0pt}}
%
%\begin{subfigure}{.5\textwidth}
%  \centering
%  \includegraphics[width=0.24\linewidth]{images/nyu_crop.png}
%  \caption{1a}
%  \label{fig:sfig1}
%\end{subfigure}%
%
%\begin{subfigure}{.5\textwidth}
%  \centering
%  \includegraphics[width=0.24\linewidth]{images/nyu_crop.png}
%  \caption{1a}
%  \label{fig:sfig1}
%\end{subfigure}%
%
%\begin{subfigure}{.5\textwidth}
%  \centering
%  \includegraphics[width=0.24\linewidth]{images/nyu_crop.png}
%  \caption{1a}
%  \label{fig:sfig1}
%\end{subfigure}%
%
%%\includegraphics[width=0.24\linewidth]{images/nyu_crop.png} \;\;\;
%%\includegraphics[width=0.33\linewidth]{images/icvl_.png} 
%%\includegraphics[width=0.32\linewidth]{images/msra1_.png} 
%\end{center}
%   \caption{Pre-processed depth images and labels from {\it HandSet}. a) NYU b) ICVL c) MSRA-2015}
%\label{fig:short}
%\end{figure}
%------------------------------------------------------------------------
\section{Related Work}
\label{sec:related}
Comprehensive reviews of hand pose estimation methods using depth sensors have been reported in \cite{barsoum2016articulated, supancic2015depth, erol2007vision}. Our work is related to the hybrid methods and the real hand pose datasets from frontal camera view. Hence, we focus on the most related works in the following subsections.
%Recently proposed CNN-based methods \cite{ge2016robust,guo2017region} have shown significant improvment in accuracy compared to hybrid methods for hand pose estimation.
%We do not claim to exceed the accuracy of recently proposed CNN based methods \cite{ge2016robust,guo2017region}, in contrast, we introduce a novel hybrid method to estimate both 3D hand pose and bone-lengths of hand skeleton from single depth frame and prove its performance comparable to the state-of-the-art hybrid methods. Furthermore, we combine available real hand pose datasets from frontal camera view.
%-------------------------------------------------------------------------
\subsection{Hand Pose Datasets Based on \textbf{\textit{Real}} Depth Data} \label{related1}
In this subsection, we briefly introduce the most commonly used real hand pose datasets. 
 
NYU hand pose dataset \cite{tompson2014real} provides $72,757$ RGB-D frames acquired from Prime Sense Carmine-1.09 depth camera. The test set contains $8252$ images. The dataset covers a wide range of complex hand poses. To acquire the ground truth, direct search method proposed by \cite{oikonomidis2011efficient} is adopted with modifications and is quite accurate. However, this dataset has no variation in hand shapes and sizes because it has only one subject in the training set and two subjects in the test set. 

ICVL dataset \cite{tang2014latent} contains $22K$ original depth frames including $10$ subjects and two test sets with $800$ frames each. However, by applying rotations, the total size of dataset exceeds $300K$ images along-with the ground truth. Intel creative gesture camera was used to acquire the depth images. The dataset has good number of complex hand poses but, not as complex as NYU dataset \cite{barsoum2016articulated}. Ground truth is created using a search method, guided by a Binary Latent Tree Model (LTM) \cite{choi2011learning}. However, ground truth is not very accurate and the variation in hand shapes and sizes is less.   

MSRA-$2015$ dataset \cite{sun2015cascaded} contains $76,500$ depth frames captured from Creative gesture camera. Images are captured from $9$ subjects, each performing $17$ hand gestures. Ground truth is annotated using a semi-automatic and iterative process followed by manual corrections \cite{qian2014realtime}. However, annotations are less accurate.  

In order to benefit from the individual pros of the above described datasets and to add the advantages of having a bigger dataset with more variations in hand shapes, sizes and type of depth cameras, we propose to combine them into one unified format described in Section \ref{sec:combi}. 

There are some other existing real hand pose datasets from frontal camera view i.e. Dexter \cite{sridhar2013interactive}, SHREC-2017\footnote {http://www-rech.telecom-lille.fr/shrec2017-hand/}, MSRA-2014 \cite{qian2014realtime}, ASTAR \cite{xu2016estimate}. However, these datasets either contain small number of original images, missing depth information, few ground truth joint positions or many outliers in the annotations. Therefore, they are not considered in this work. 
%SHREC-$2017$ dataset\footnote {http://www-rech.telecom-lille.fr/shrec2017-hand/} for hand-poses offers $164K$ depth frames captured from Intel Creature Gesture Camera. Number of participating users are $28$ and each of them was asked to do $14$ hand gestures. This dataset has largest number of original frames to date, however, the ground truth is highly inaccurate for huge number of data frames. For this reason, we do not consider this dataset in our work.
%   \begin{figure}[t]
%     \subfloat[\label{subfig-1:dummy}]{%
%       \includegraphics[width=0.12\textwidth]{images/nyu_.png}
%     }
%     \;\;\;
%     \subfloat[\label{subfig-2:dummy}]{%
%       \includegraphics[width=0.16\textwidth]{images/icvl_.png}
%     }
%     \hfill
%     \subfloat[\label{subfig-2:dummy}]{%
%       \includegraphics[width=0.16\textwidth]{images/msra_.png}
%     }
%     \caption{Pre-processed depth images and labels from {\it HandSet}. Sample images from a) NYU b) ICVL c) MSRA-2015}
%     \label{fig:datasets}
%   \end{figure}
%-------------------------------------------------------------------------
\begin{figure}[t]
\begin{center}
% \fbox{\rule{0pt}{2in} \rule{0.9\linewidth}{0pt}}
   \includegraphics[width=0.6\linewidth]{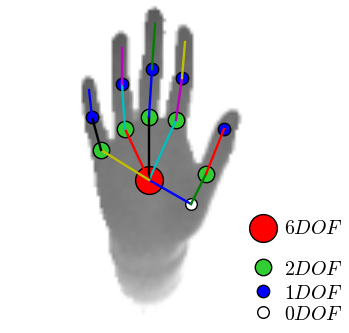}
\end{center}
   \caption{Illustration of our hand skeleton with 21DOF.}
\label{fig:long}
\vspace{-3mm}
\end{figure}

\subsection{Hybrid Methods for Hand Pose Estimation}

The first CNN-based hand pose estimation method was introduced by \cite{tompson2014real}. Joint locations are predicted from CNN in the form of heatmaps. Thereafter, an inverse Kinematics (IK) is applied to estimate 3D hand pose based on predicted joints. Poier et al. \cite{poier2015hybrid} use a model based optimization step based on  multiple 3D joint hypothesis (proposal distributions) received from a random regressor. In \cite{sridhar2015fast}, coarse joints are predicted using pixel classification random forest algorithm. In the generative model fitting step, a similarity function is optimized between the predicted joints and generated joints. In the methods mentioned above, model fitting (generative) is separated from the joints estimation part. In \cite{oberweger2015training}, Obreweger et al. perform a complex training of a feedback loop to infer the correct hand pose. It uses three neural networks. First, to estimate coarse hand pose. Second, is used to synthesize the input image. Third, comprises of pose update network. Ye et al. \cite{ye2016spatial} introduce a hierarchical hybrid method with a spatial attention mechanism and hierarchical Particle Swarm Optimization (PSO). Zhou et al. \cite{zhou2016model} propose a low latency framework that seamlessly integrates a generative hand model layer with  a neural network. A generative hand model layer is introduced to map the received joint angles to 3D positions. However, the hand model requires to be calibrated for a specific user. Inspired by this work, we propose a new low latency hybrid algorithm for estimating hand skeleton bone-lengths and pose simultaneously. The end-to-end training of our pipeline is simple and highly efficient. The forward kinematic function in the generative layer is differentiable with respect to joint angles and hand scale parameters. 
%  Recently, \cite{mueller2017real} uses direct joint regression and heatmaps to estimate 3D joint positions. A kinematic pose fitting energy further refines the hand pose. However, synthetic data is used for training and networks are complex. 
%In \cite{sridhar2016real}, Sridhar et al. proposes a more accurate and fast hand tracking algorithm. An objective function is optimized, between the part labels received from per-pixel random forest classifier and a Gaussian Mixture Model (GMM) representation of input depth frame, to estimate the hand pose. Taylor et al.\cite{taylor2016efficient}, proposes a non-linear discriminative objective function optimization scheme by using gradients provided by a smooth surface model. The proposed model fitting function is costly but requires less number of iterations.
%-------------------------------------------------------------------------
%  Samples of pre-processed images from three datasets along with their ground truth annotations are shown in Figure \ref{fig:datasets}. 
\vspace{-1mm}
\section{Combined Dataset and Pre-Processing}
\label{sec:combi}
First step to merge different datasets is to select the number of common joint positions present in all datasets. ICVL dataset has least number of joints. We consider corresponding 16 joints in the NYU and MSRA-$2015$ datasets and remove additional joints for consistency. Since, each dataset uses different depth camera to acquire images, we need to pre-process the depth frames according to their respective camera intrinsics, frame resolutions and depth range. Inspired by the method in \cite{zhou2016model}, for depth invariance, the images are cropped around palm center in all three dimensions (u, v and depth) using a fixed size bounding box. Then, depth values are normalized to $[$$-1$, $1$$]$. The 3D joint locations are also normalized in range $[$$-1$, $1$$]$ using the bounding box. The final pre-processed image is of 128 x 128 dimension and has $16$ ground truth annotations which include $12$ internal joints as shown in Figure \ref{fig:long} and four finger-tips. The {\it HandSet} contains $450K$ pre-processed training depth images, $18K$ test images and $20$ different subjects.   

\begin{figure}[t]
\begin{center}
% \fbox{\rule{0pt}{2in} \rule{0.9\linewidth}{0pt}}
   \includegraphics[width=0.9\linewidth]{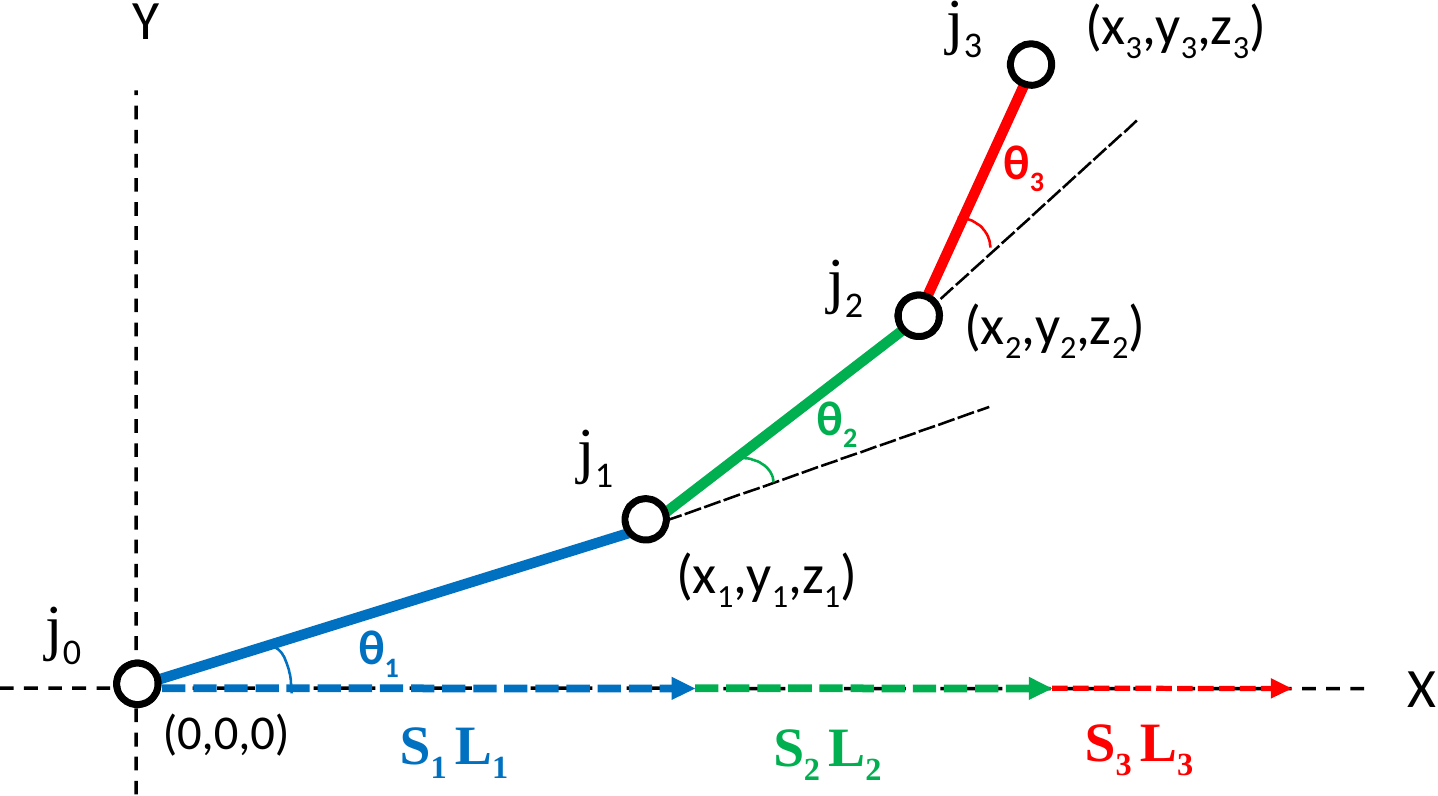}
\end{center}
   \vspace{-2mm}
   \caption{Illustration of 3D joint transformations of four adjacent joints of initial hand skeleton using forward kinematics process. Assuming 1DOF for each joint and considering three rotations at $j_0,j_1,j_2$ among $z$ axis, the relative position of joint $j_3$  with respect to reference joint position $j_0$ can be calulated as $(x_3,y_3,z_3,1)^T = [\textrm {Trans}_x({\bf S_1}L_1)]\times[\textrm {Rot}_z(\theta_1)]\times[\textrm {Trans}_x({\bf S_2}L_2)]\times[\textrm {Rot}_z(\theta_2)]\times[\textrm {Trans}_x({\bf S_3}L_3)]\times[\textrm {Rot}_z(\theta_3)]\times[0,0,0,1]^T.$ }
   % new line in equation $\\$
\label{fig:FK}
\vspace{-5mm}
\end{figure}

\begin{figure*}[t]
   \centering    
     \subfloat{%
       \includegraphics[width=0.95\textwidth]{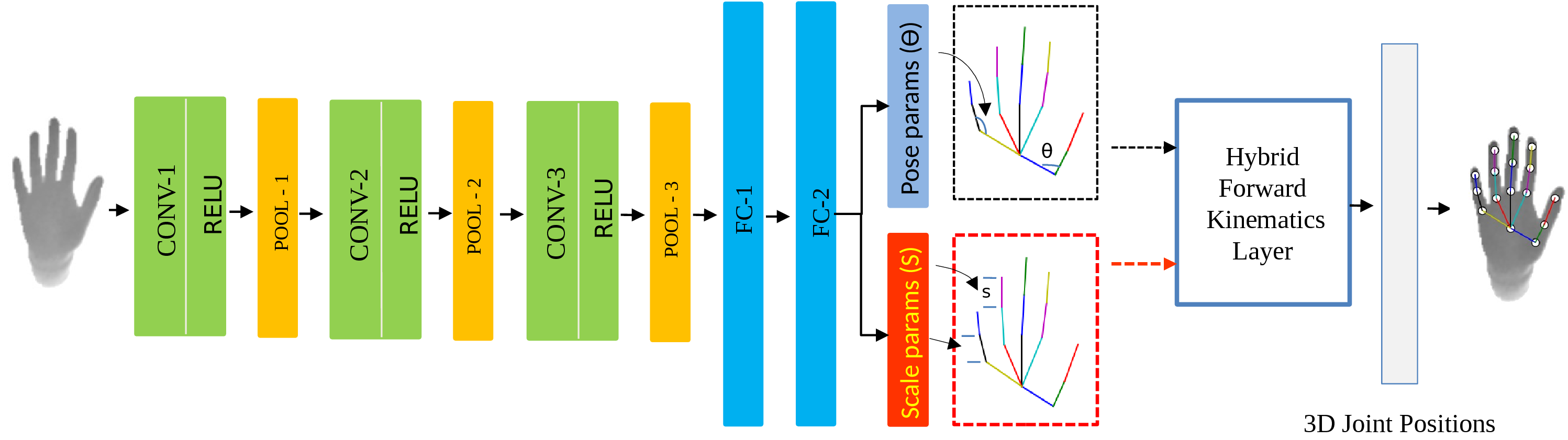}} % figure_4-crop.pdf
   \vspace{-2mm}
   \caption{Illustration of our model for simultaneous hand pose and skeleton estimation. The algorithm starts from three convolutional layers and two fully connected layers. The last fully connected layer outputs hand pose parameters ($\Theta$) and scale parameters ($S$) associated with the bone-lengths of the skeleton. In the end, a hybrid forward kinematic function is applied that outputs 3D joint positions using the hand scale and pose parameters.}
\label{fig:Deep Scale}
\vspace{-3mm}
\end{figure*}
    
%-------------------------------------------------------------------------
\vspace{-2mm}
\section{Hand Pose and Bone-Lengths Estimation}
In this section, we explain our approach for simultaneous estimation of hand pose and bone-lengths of the hand skeleton using a hybrid forward kinematics layer and deep architectures.
%-------------------------------------------------------------------------
% \vspace{5mm}
\subsection{Hybrid Forward Kinematics Layer}
\label{sec:hyb}
%The skeleton has $21$ degrees of freedom (DoF) for hand pose parameters defined on $16$ hand joints. Six DoFs are reserved for global palm orientation and position. All remaining DoFs are defined for the rest of hand joints. We assume zero pose vector as the reference hand pose shown in Figure \ref{fig:long}. All other poses are defined relative to this reference hand pose.
Figure \ref{fig:long} shows our hand skeleton. We assume a zero pose vector (i.e. pose with all parameters set to zero) as the reference hand pose. All other poses are defined relative to this reference pose. We initialize the hand skeleton by the averages of individual bone-lengths from ground truth annotations of each dataset. Given the hand pose and scale parameters, the hybrid forward kinematic layer (see Figure \ref{fig:Deep Scale}) implements a forward kinematic function~$\textit{F}_k$  defined as:
\begin{equation} \label{eq:1}
\textit{F}_k(\Theta,S) = \textit{J} 
\end{equation}
Where~$\Theta= \left \{\theta_p \right \}$,~$p = \left \{1,2,\cdots,21 \right \}$ is a vector of pose parameters,~$S= \left \{s_l \right \}$,~$l = \left \{1,2,\cdots,15 \right \}$ defines the hand scale factors associated with bone-lengths and~$\textit{J} = \left \{j_n \right \}$,~$n = \left \{1,2,\cdots,16 \right \}$  is a vector of the predicted joint positions. 

The 3D transformation of each of the $16$ joints in~$\textit{J}$ is derived from its joint angles for rotation and scaled bone-lengths for translation. The global 3D position ~$(x_n,y_n,z_n)$ of a joint is obtained by applying series of transformations (rotational and translational) along the path starting from hand root joint to this joint as shown in Figure \ref{fig:FK}.

Cost function is obtained by using Euclidean 3D joint location loss given as:
\begin{equation} \label{eq:2}
\frac{1}{2}\norm{ \textit{F}_k(\Theta,S) - {\textit{J}_G}_T }^2
\end{equation}
Where ${\textit{J}_G}_T$ is a vector of 3D ground truth joint positions.

Since, Equation \ref{eq:1} is differentiable with respect to both pose parameters $\Theta$ and hand scales $S$, hence, it can be used in deep network to compute gradients for back-propagation.
% The Jacobians of $\textit{F}_k$ with respect to $\Theta$ and $S$ are:
The Jacobian of $\textit{F}_k$ with respect to $\Theta$ is defined as:
\begin{equation} \label{eq:3}
\frac{\partial \textit{F}_k}{\partial \Theta} = 
\begin{bmatrix} {\partial j_1 \over \partial \theta_1} & {\partial j_1 \over \partial \theta_2} & \cdots & {\partial j_1 \over \partial {\theta_2}_1} \\

{\partial j_2 \over \partial \theta_1} & {\partial j_2 \over \partial \theta_2} & \cdots & {\partial j_2 \over \partial {\theta_2}_1}  \\

\vdots & \vdots & \ddots & \vdots \\

{\partial {j_1}_6 \over \partial \theta_1} & {\partial {j_1}_6 \over \partial \theta_2} & \cdots & {\partial {j_1}_6 \over \partial {\theta_2}_1} 
 \end{bmatrix}
\end{equation}

%\begin{equation} \label{eq:4}
%\frac{\partial \textit{F}_k}{\partial S} = 
%\begin{bmatrix} {\partial j_1 \over \partial s_1} & {\partial j_1 \over \partial s_2} & \cdots & {\partial j_1 \over \partial {s_1}_5} \\
%
%{\partial j_2 \over \partial s_1} & {\partial j_2 \over \partial s_2} & \cdots & {\partial j_2 \over \partial {s_1}_5}  \\
%
%\vdots & \vdots & \ddots & \vdots \\
%
%{\partial {j_1}_6 \over \partial s_1} & {\partial {j_1}_6 \over \partial s_2} & \cdots & {\partial {j_1}_6 \over \partial {s_1}_5} 
% \end{bmatrix}
%\end{equation}
The Jacobian of $\textit{F}_k$ with respect to $S$ can be defined in a similar way. 
Partial derivative of a joint $j_n$ in ${\it J}$ with respect to a pose parameter $\theta_p$ can be calculated as:
\begin{equation} \label{eq:5}
\frac{\partial \textit{j}_n}{\partial \theta_p} = 
\big(\prod_{c \in P_c} [\textrm {Rot}_\phi(\theta)] \times [{\textrm {Trans}_\phi}_c(S_cL_c)]\big)[0,0,0,1]^T 
\end{equation}
where,
$$
\textrm {Rot}_\phi(\theta) = \left\{ \begin{array}{rl}
 {\textrm {Rot}_\phi}_c(\theta_c) &\mbox{ if $c \neq p$} \\\\
  {\textrm {Rot}_\phi}_c'(\theta_c) &\mbox{ if $c = p$}
       \end{array} \right.
$$
% $P_c$ is the set of parent joints of $j_n$ connected in path to the root joint and $\phi$ is the rotation axis.
$P_c$ is the set of joints along kinematic chain from $j_n$ to the root joint and $\phi$ is the rotation axis.

Similarly, we compute partial derivative of a joint $j_n$ in ${\it J}$ with respect to a scale parameter $s_l$ as:
\begin{equation} \label{eq:6}
\frac{\partial \textit{j}_n}{\partial s_l} = 
\sum_{k \in P_k} \big[\big(\prod_{c \in P_c} [{\textrm {Rot}_\phi}_c(\theta_c)] \times [\textrm {Trans}_\phi(SL)]\big)[0,0,0,1]^T \big] 
\end{equation}
where,
$$
\textrm {Trans}_\phi(SL) = \left\{ \begin{array}{rl}
 {\textrm {Trans}_\phi}_c(S_cL_c) &\mbox{ if $c \neq k$} \\\\
  {\textrm {Trans}_\phi}_c'(S_cL_c) &\mbox{ if $c = k$}
       \end{array} \right.
$$
and, $P_k$ is the set of parent joints of $j_n$ that share the same scale parameter $s_l$. 

%calculated by mulplications of the derivative of its rotation matrix and all other transformation matrices of its connected joints to the root joint (palm center). 

   \begin{figure*}[t]
     \rotatebox{90}{\;\;\; \small NYU Dataset}
     \subfloat{\includegraphics[width=0.14\textwidth]{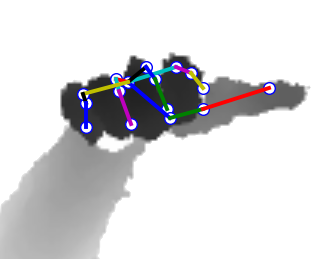}
     } \;\;\;
     \subfloat{\includegraphics[width=0.14\textwidth]{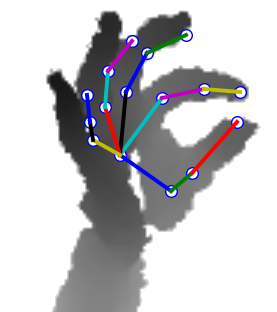}
     } \;\;\;
     \subfloat{\includegraphics[width=0.14\textwidth]{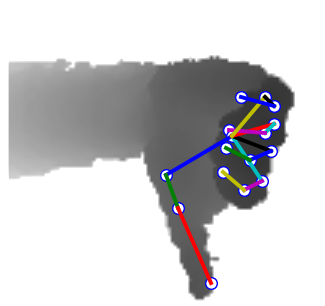}
     } \;\;\;
     \subfloat{\includegraphics[width=0.14\textwidth]{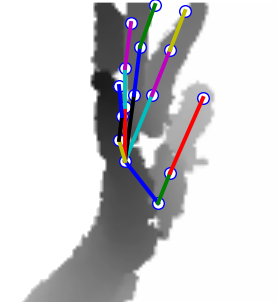}
     } \;\;\;
     \subfloat{\includegraphics[width=0.14\textwidth]{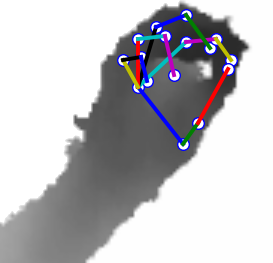}
     } \hfill
     \subfloat{\includegraphics[width=0.14\textwidth]{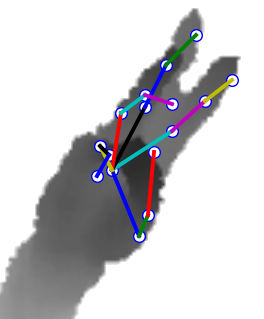}
     }
     
     \rotatebox{90}{\small MSRA-2015 Dataset}
     \subfloat{\includegraphics[width=0.14\textwidth]{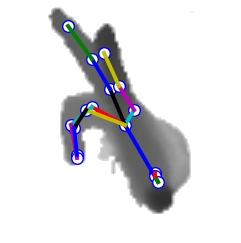}
     } \;\;\;
     \subfloat{\includegraphics[width=0.14\textwidth]{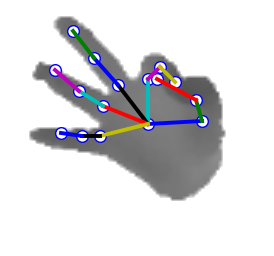}
     } \;\;\;
     \subfloat{\includegraphics[width=0.14\textwidth]{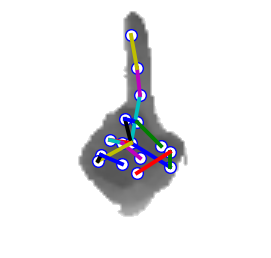}
     } \;\;\;
     \subfloat{\includegraphics[width=0.14\textwidth]{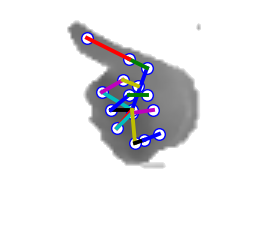}
     } \;\;\;
     \subfloat{\includegraphics[width=0.14\textwidth]{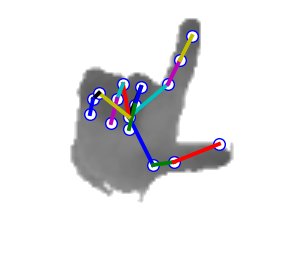}
     } \hfill
     \subfloat{\includegraphics[width=0.14\textwidth]{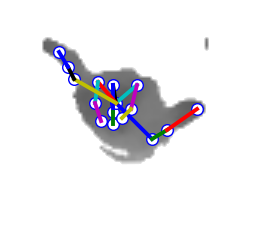}
     }
     
     \rotatebox{90}{\;\;\; \small ICVL Dataset}
     \subfloat{\includegraphics[width=0.14\textwidth]{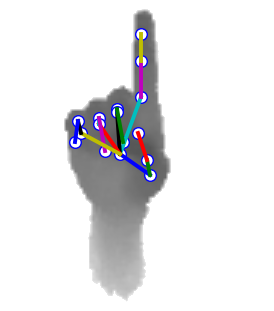}
     } \;\;\;
     \subfloat{\includegraphics[width=0.14\textwidth]{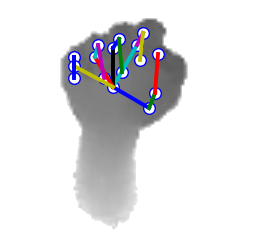}
     } \;\;\;
     \subfloat{\includegraphics[width=0.14\textwidth]{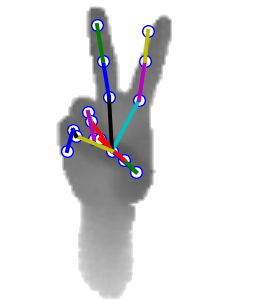}
     } \;\;\;
     \subfloat{\includegraphics[width=0.14\textwidth]{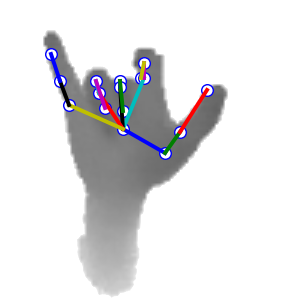}
     } \;\;\;
     \subfloat{\includegraphics[width=0.14\textwidth]{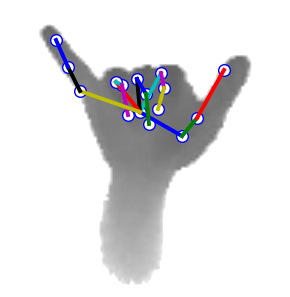}
     } \hfill
     \subfloat{\includegraphics[width=0.14\textwidth]{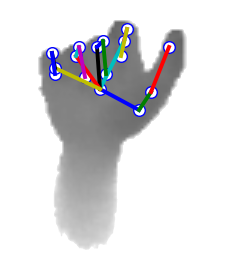}
     }

     \caption{Sample results from our {\bf 5Scales} architecture. The predicted 3D joint positions are displayed on the depth images. The rows show images from NYU, MSRA-2015 and ICVL datasets, respectively from top to bottom. }
     \label{fig:5scales_final}
     \vspace{-2mm}
   \end{figure*}

   \begin{figure}[!ht]
   
     \subfloat{%
     \rotatebox{90}{\; \small Our Model}
       \includegraphics[width=0.09\textwidth]{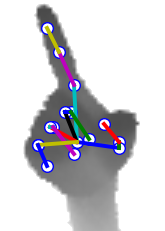}
     }
     \;\;\;\;\;\;\;\;\;\;
     \subfloat{%
       \includegraphics[width=0.11\textwidth]{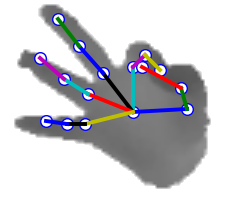}
     }
     \;\;\;\;\;\;\;\;\;\;
     \subfloat{%
       \includegraphics[width=0.10\textwidth]{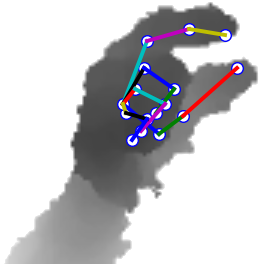}
     }
     
     \subfloat{%
     \rotatebox{90}{\; \small Zhou et al.\cite{zhou2016model}}
       \includegraphics[width=0.09\textwidth]{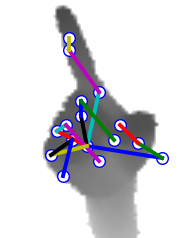}
     }
     \;\;\;\;\;\;\;\;\;\;
     \subfloat{%
       \includegraphics[width=0.11\textwidth]{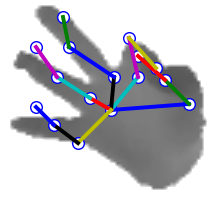}
     }
     \;\;\;\;\;\;\;\;\;\;
     \subfloat{%
       \includegraphics[width=0.10\textwidth]{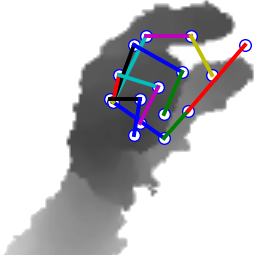}
     }
     
     \hfill   
     ICVL \;\;\;\;\;\;\;\;\;\;\;\;\;\;  MSRA-2015 \;\;\;\;\;\;\;\;\;\;\;\;  NYU  \;\;\;\;\;\;\;\;\;\;
     \caption{Sample images with overlaid predicted 3D joint positions from our model with hand scale parameters (top row) and Zhou et al. \cite{zhou2016model} without hand scale parameters (bottom row), when trained on {\it HandSet} dataset.}
     \label{fig:wscale_woscale}
     \vspace{-3mm}
   \end{figure} 
   
\subsection{Deep Architectures with Hand Scales}

Human hands differ in individual fingers and palm sizes. There is a need to explicitly consider such differences during training. Therefore, we introduce various scales of hand as additional learning parameters to facilitate CNN training on {\it HandSet} as shown in Figure \ref{fig:Deep Scale}. These scales factors are learned by the CNN along-with the pose parameters.

We propose three implementations of our method explained in the following subsections and compare their performances in Section \ref{sec:results}. We build our CNN architecture based on the baseline architecture proposed in \cite{oberweger2015hands}, mainly for the sake of fair comparison. The pipeline of our algorithm is shown in Figure \ref{fig:Deep Scale}. The architecture of CNN comprises of 3 convolutional layers using 5, 5, 3 kernel sizes respectively. Max pooling layers are then connected using strides 4,2,1 with zero padding. The feature maps from convolutional layers are of size 12 x 12 x 8.  Two fully connected layers consist of 1024 neurons each. Dropout layers are added with dropout ratio of 0.3. All convolutional layers use ReLu as activation.       
% We tried several CNN architectures for learning hand pose and scale parameters. However, we found that the baseline architecture proposed in \cite{oberweger2015hands} works best in our case. Therefore, w
%\vspace{-3.5mm}
\subsubsection{GlobalScale}
In this architecture, we define a global scale for the hand skeleton such that it can symmetrically vary its size. In Figure \ref{fig:Deep Scale}, the last fully connected layer outputs pose parameters and additional global hand scale parameter $s$, shared by all $15$ bones of the hand skeleton. Larger scale value results in bigger hand skeleton and vice versa. The hybrid forward kinematic layer takes this scale parameter as input along-with pose parameters and computes 3D joint positions according to Equation \ref{eq:1}. The partial derivative of a joint with respect to the global scale parameter can be computed using Equation \ref{eq:6}.  

% The respective gradients of global scale and pose parameters are propagated back to the CNN. The Jacobian matrix given by Equation \ref{eq:4} can be re-defined as:
%\begin{equation} \label{eq:7}
%\frac{\partial \textit{F}_k}{\partial s} = 
%\begin{bmatrix} {\partial j_1 \over \partial s} & {\partial j_2 \over \partial s} & \cdots & {\partial {j_1}_6 \over \partial s}
% \end{bmatrix}^T
%\end{equation}
%\vspace{-3.5mm}
\subsubsection{5Scales}
This architecture associates five separate hand scale parameters from tips of the five fingers to the palm center (root joint). These parameters allow the individual fingers to vary their lengths according to their respective scale values, thereby adding a flexibility to both shape and size of the hand skeleton. These parameters are defined by $S$ as:
\begin{equation} \label{eq:8}
S= \left \{{s_f}_1,{s_f}_2,{s_f}_3,{s_f}_4,{s_f}_5 \right \}
\end{equation}
Given the pose parameters $\Theta$  and $S$, forward kinematic function defined by Equation \ref{eq:1} is applied to estimate more accurate 3D joint locations. Using Equation \ref{eq:6}, the partial derivative of a joint with respect to its associated finger scale parameter is calculated.  

%The Jacobian matrix defined by Equation \ref{eq:4} is modified as:
%
%\begin{equation} \label{eq:9}
%\frac{\partial \textit{F}_k}{\partial S} = 
%\begin{bmatrix} {\partial j_1 \over \partial {s_f}_1} & {\partial j_1 \over \partial {s_f}_2} & \cdots & {\partial j_1 \over \partial {s_f}_5} \\
%
%{\partial j_2 \over \partial {s_f}_1} & {\partial j_2 \over \partial {s_f}_2} & \cdots & {\partial j_2 \over \partial {s_f}_5}  \\
%
%\vdots & \vdots & \ddots & \vdots \\
%
%{\partial {j_1}_6 \over \partial {s_f}_1} & {\partial {j_1}_6 \over \partial {s_f}_2} & \cdots & {\partial {j_1}_6 \over \partial {s_f}_5} 
% \end{bmatrix}
%\end{equation}
\subsubsection{MultiScale}
In this architecture, we assign a separate scale to each bone of our hand skeleton. Each bone-length can be estimated independently of other bones (see Section \ref{sec:hyb}). Hence, this architecture provides the maximum flexibility to adapt shape and size of the hand skeleton. 

%The set S used in hybrid forward kinematics layer can be written as:
%\begin{equation} \label{eq:10}
%S= \left \{s_1,s_2, \cdots  , {s_1}_5 \right \}
%\end{equation}
%The corresponding gradients are propagated back to the CNN according to Equation \ref{eq:3} and \ref{eq:4}. 
\section{Implementation Details}
For end-to-end training of our model, we use Caffe open source framework for deep networks \cite{jia2014caffe}. The network is trained until convergence with a fixed learning rate of $0.001$  using $0.9$ as SGD momentum. We perform data augmentations i.e. rotations and scalings during training phase. The complete framework runs on a PC with Nvidia GeForce 1070 GPU. One forward pass takes $7ms$.

   \begin{figure}[t]  
   \centering    
     \subfloat{%
       \rotatebox{90}{\;\;\;\;\;\;\;\; \small USER 1}
       \includegraphics[width=0.15\textwidth]{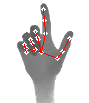}
     }
     \subfloat{%
       \includegraphics[width=0.15\textwidth]{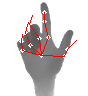}
     }
     \subfloat{%
       \includegraphics[width=0.16\textwidth]{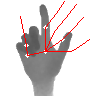}}

     \subfloat{%
     \rotatebox{90}{\;\;\;\;\;\;\;\; \small USER 2}
       \includegraphics[width=0.15\textwidth]{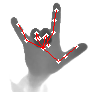}
     }
     \subfloat{%
       \includegraphics[width=0.15\textwidth]{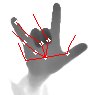}
     }
     \subfloat{%
       \includegraphics[width=0.15\textwidth]{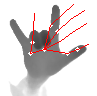}}
     
     \subfloat{%
      \rotatebox{90}{\;\;\;\;\;\;\;\; \small USER 3}
       \includegraphics[width=0.16\textwidth]{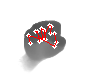}
     }
     \subfloat{%
       \includegraphics[width=0.14\textwidth]{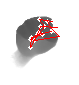}
     }
     \subfloat{%
       \includegraphics[width=0.16\textwidth]{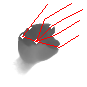}}

     \hfill   
     Our Model \;\;\;\;\;\;  Zhou et al.\cite{zhou2016model} \;\;\;\;\;  Zhou et al.\cite{zhou2016model}  \;\;\;\;
     
     \hfill   
     ( {\it HandSet} ) \;\;\;\;\;\;\;\;  ( {\it HandSet} ) \;\;\;\;\;\;\;\;  ( NYU only )  \;\;\;\;\;\;
     \caption{Hand pose inference results on unseen images from our model and Zhou et al. \cite{zhou2016model}. Our model shows good results while the compared model fails to converge.}
     \label{fig:general_hands}
     \vspace{-2mm}
   \end{figure}

   \begin{figure}
     \subfloat[]{%
       \includegraphics[width=0.109\textwidth]{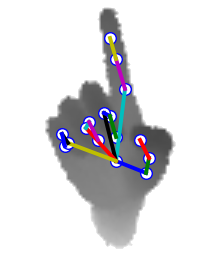}
     }
     \subfloat[]{%
       \includegraphics[width=0.13\textwidth]{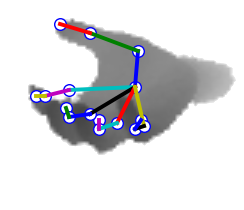}
     }
     \subfloat[]{%
       \includegraphics[width=0.11\textwidth]{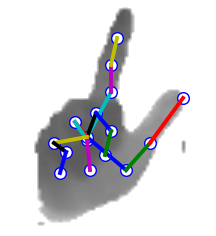}}
     \;
     \subfloat[]{%
       \includegraphics[width=0.103\textwidth]{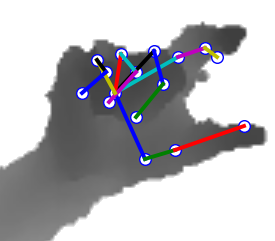}
     }
     \caption{Some failure cases are shown from our {\bf GlobalScale} architecture (a and b) and {\bf MultiScale} architecture (c and d). In (a) and (b), smaller global scale leads to incorrect poses. In (c) and (d), we see inconsistency in bone-lengths due to independent scales estimation.}
     \label{fig:multi-global}
     \vspace{-3mm}
   \end{figure}
% xu2017lie     
\section{Results}
\label{sec:results}
In this section, we illustrate the accuracy of our model through both qualitative and quantitative results and comparisons with the state-of-the-art hybrid methods. We do not claim to exceed the accuracy of recently published discriminative methods \cite{ge2017robust,guo2017region} which neglect hand model geometry i.e. kinematics and physical constraints. Instead, we provide a performance comparison with the existing hybrid methods to validate our algorithm that fully exploits a flexible hand model geometry and estimates the 3D hand pose and bone-lengths of the hand skeleton simultaneously. Notably, famous public datasets such as NYU and ICVL contain low variation in hand shapes and sizes (see Section \ref{related1}). However, we demonstrate our results on these datasets for completeness. We use two common evaluation metrics. First is the average 3D joint location error on test dataset. Second, fraction of test frames for which maximum predicted 3D joint error is below a certain threshold in millimeter.

%   \begin{figure*}[t]  
%   \centering    
%     \subfloat{%
%       \includegraphics[width=5.7cm,height=4.45cm]{images/Quantitative_Results/frame_error_all.png}}
%     \;
%     \subfloat{%
%       \includegraphics[width=5.5cm,height=4.45cm]{images/Quantitative_Results/frame_error_nyu.png}}
%     \;
%     \subfloat{%
%       \includegraphics[width=5.9cm,height=4.5cm]{images/Quantitative_Results/frame_error_icvl.png}}
%       
%     \subfloat{%
%       \includegraphics[width=5.7cm,height=4.45cm]{images/Quantitative_Results/joints_error.png}}
%     \;
%     \subfloat{%
%       \includegraphics[width=5.5cm,height=4.4cm]{images/Quantitative_Results/joints_error_nyu_5scales1.png}}
%     \;
%     \subfloat{%
%       \includegraphics[width=5.9cm,height=4.42cm]{images/Quantitative_Results/joints_error_icvl.png}}
%  
%     \hfill   
%     (a) {\it HandSet} dataset\;\;\;\;\;\;\;\;\;\;\;\;\;\;\;\;\;\;\;\;\;\;\;\;\;\;\;\;\;\; (b) NYU dataset \;\;\;\;\;\;\;\;\;\;\;\;\;\;\;\;\;\;\;\;\;\;\;\;\;\;\;\;\;\;\;\;\;\;\;   (c) ICVL dataset \;\;\;\;\;\;\;\;\;\;\;\;\;\;\;\;\;
%     \vspace{-2mm}
%     \caption{ Comparison of our architectures and with baseline hybrid \cite{zhou2016model} on {\it HandSet} is shown in (a). Comparison on public datasets with baseline hybrid methods is given in (b) and (c). Percentage of success frames (upper) and mean error (lower). } 
%     \label{fig:Quantitative_2}
%   \end{figure*}
   
\subsection{Qualitative Evaluation}
%by inspection, We found that 5Scales architecture performs the best as compared to our other two models (GlobalScale and MultiScale). 
%The qualitative results from our three models (GlobalScale, MultiScale and 5Scales) and the state-of-the-art hybrid method \cite{zhou2016model} on {\it HandSet} are shown in Figure \ref{fig:Quantitative_2}a. 
%Our 5Scales model outperforms the deep model approach in \cite{zhou2016model} and also, proves to be the best among our three models. Some failure cases from our two other architectures (GlobalScale and MultiScale) are shown in Figure \ref{fig:multi-global}. 
Some challenging hand pose images from three datasets along-with predicted joint positions from our model are shown in Figure \ref{fig:5scales_final}. We show some sample images with overlaid hand skeleton from our 5Scales model and deep model \cite{zhou2016model} in Figure \ref{fig:wscale_woscale}. Our model shows very good results whereas, the compared model is unable to converge successfully leading to inaccurate 3D hand joint positions and bone-lengths. We tested the 5Scales model with Zhou et al. \cite{zhou2016model} on unseen images acquired from three different users. Our model is able to infer hand pose quite accurately whereas, the other model fails to converge (see Figure \ref{fig:general_hands}). Some failure cases from our two other architectures (GlobalScale and MultiScale) are shown in Figure \ref{fig:multi-global}. Incorrect bone-lengths estimation from GlobalScale architecture can happen due to a single scale parameter associated with all bones of the hand skeleton. On the other hand, in MultiScale architecture, independent learning of each bone-length of the hand skeleton may result in incorrect bone-lengths estimation.

\begin{table}
\begin{center}
\begin{tabular}{|l|c|}
\hline
Methods & 3D Joint Location Error \\
\hline\hline
Zhou et al. \cite{zhou2016model} & 18.7mm \\
MultiScale [Ours] & 15.1mm \\
GlobalScale [Ours] & 15.3mm \\
{\bf 5Scales} [Ours] & {\bf 12.7mm} \\
\hline
\end{tabular}
\end{center}
\vspace{-5mm}
\caption{Quantitative comparison of our three architectures and Zhou et al. \cite{zhou2016model} on {\it HandSet} test dataset.}
\label{tab:Table_1}
\end{table}

\begin{table}
\begin{center}
\begin{tabular}{|l|c|}
\hline
Methods & 3D Joint Location Error \\
\hline\hline
% Oberweger et al. Prior.\cite{oberweger2015hands} & 21.0mm \\
Oberweger et al. \cite{oberweger2015training} & \bf{16.0mm} \\
Zhou et al. \cite{zhou2016model} & 17.0mm \\
{\bf Ours} & {16.2mm}\\
\hline
\end{tabular}
\end{center}
\vspace{-5mm}
\caption{Quantitative comparison on {\bf NYU} test set.}
\label{tab:Table_2}
\end{table} 

\begin{table}
\begin{center}
\begin{tabular}{|l|c|}
\hline
Methods & 3D Joint Location Error \\
\hline\hline
% Oberweger et al. Prior.\cite{oberweger2015hands} & 21.0mm \\
LRF \cite{tang2014latent} & 12.6mm \\
Zhou et al. \cite{zhou2016model} & 11.5mm \\
{\bf Ours} & {\bf 10.0mm}\\
\hline
\end{tabular}
\end{center}
\vspace{-5mm}
\caption{Quantitative comparison on {\bf ICVL} test set.}
\label{tab:Table_3}
\vspace{-4mm}
\end{table}

\subsection{Quantitative Evaluation} 
\begin{figure}[t]  
   \centering    
     \subfloat{%
       \includegraphics[width=7.8cm,height=5.8cm]{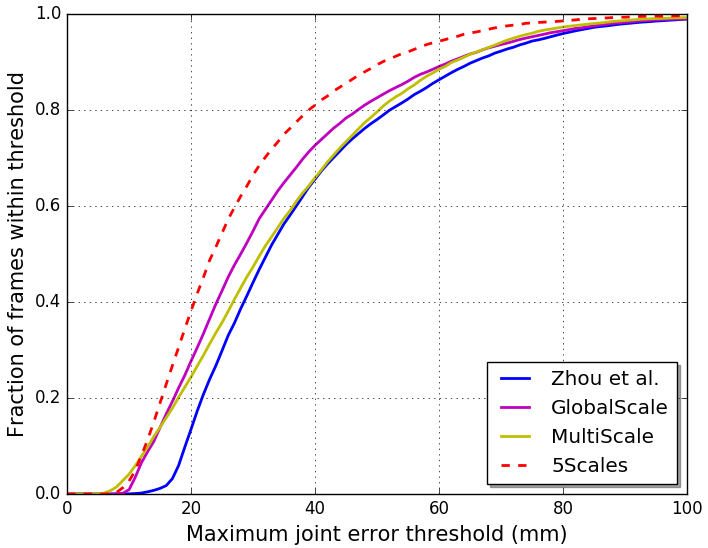}
     }
     \;
     \subfloat{%
      \includegraphics[width=7.8cm,height=5.8cm]{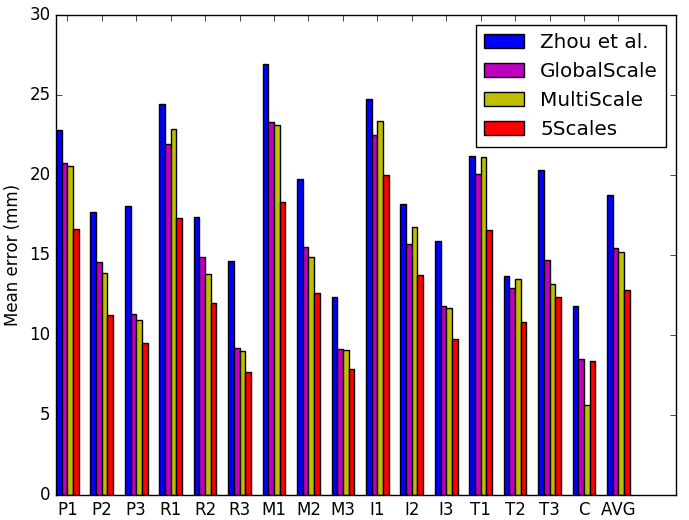}}
     \caption{Qualitative comparison of our proposed architectures (GlobalScale, MultiScale and 5Scales) vs. Zhou et al. \cite{zhou2016model} on {\it HandSet} test dataset. The upper shows the fraction of frames in error within thresholds and the lower shows the mean error on individual joints.}
     \label{fig:Quantitative_1}
     \vspace{-3mm}
   \end{figure}

\begin{figure}[t]  
   \centering    
     \subfloat{%
       \includegraphics[width=7.8cm,height=5.7cm]{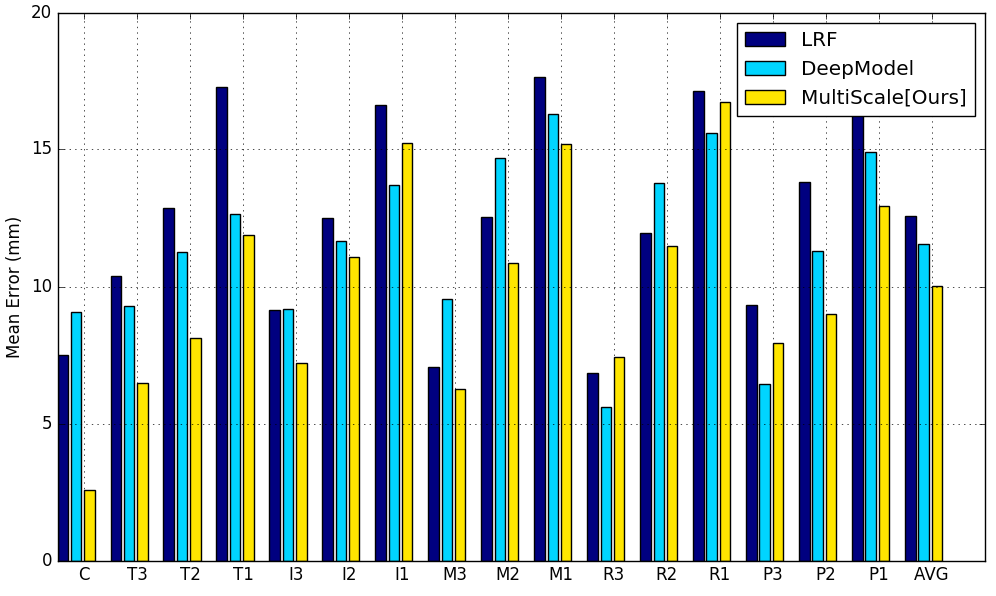}}
     \vspace{-2mm}
     \caption{ comparison with respect to mean error on individual joints with the state-of-the-art hybrid methods (LRF \cite{tang2014latent}, DeepModel \cite{zhou2016model}) on ICVL dataset.} 
     \label{fig:Quantitative_2}
     \vspace{-3mm}
   \end{figure}

We trained our three architectures (GlobalScale, MultiScale and 5Scales) as well as publicly available model based deep architecture \cite{zhou2016model} on {\it HandSet}. Notably, \cite{zhou2016model} fails when trained on {\it HandSet}. This is mainly due to the fact that they assume a fixed hand model geometry during end-to-end training. We summarize the comparison of accuracies in Figure \ref{fig:Quantitative_1} and Table \ref{tab:Table_1}. Our 5scales architecture shows the best accuracy and proves that our approach works well with large variation in hand shapes and sizes. On NYU dataset, our accuracy is comparable to Oberweger et al. \cite{oberweger2015training} on common joints (see Table \ref{tab:Table_2}). On ICVL dataset, our method shows improved performance in comparison to other state-of-the-art hybrid methods (see Table \ref{tab:Table_3}). Since, NYU dataset has no variation (one subject) and ICVL has low variation in hand shapes and sizes, therefore one can see a clear advantage of our method on ICVL dataset while a comparable performance on NYU dataset. Figure \ref{fig:Quantitative_2} shows a more detailed comparison on individual joints in ICVL dataset.  

%-------------------------------------------------------------------------
\section{Conclusion and Future Work}
In this work, we present a novel hybrid method that outputs 3D hand pose as well as bone-lengths of the hand skeleton simultaneously. We demonstrate the effectiveness of our approach on depth images captured from unseen subjects. Our method uses one CNN and a hybrid forward kinematics layer to predict 3D joint positions of the hand from a single depth image. The CNN estimates hand scale parameters (associated to bones of hand) and pose parameters. In the hybrid forward kinematics layer, the initial hand skeleton is reshaped according to estimated hand scale parameters and a differentiable forward kinematic function is applied. Three different implementations of our method are introduced that describe the hand scale parameters in distinct ways. In addition, we present a unified pre-processing method to combine famous real hand pose datasets for better training of the CNN thereby, gaining an advantage of bigger dataset with large variation in hand shapes and sizes in particular. The training process is simple and efficient and proposed algorithm is well suited for real-time applications. Qualitative and quantitative results verify that our method achieves improved performance over the state-of-the-art hybrid methods. 

This work can be further extended in some interesting dimensions. It can be combined with other outperforming discriminative methods to achieve higher accuracy. We plan to extend this work for stable real-time hand tracking. In this case, small variations in hand scale may occur for the same person. This can be addressed by automatically fixing the estimated bone-lengths after a few frames. The hand skeleton can be upgraded to skinned hand model for fine representation of hand shape and size, thereby learning more complex hand shape parameters using CNN. We further plan to enlarge the combined dataset by including more variety of hand shapes and sizes using both real and synthetic images. The same can be extended to simultaneous human body pose and shape estimation. 
% Also, physical constraints can be imposed on estimated bone-lengths during training and tracking. 
      
% Specifically, the learned bone-lengths through CNN increase flexibility in an initial hand skeleton to adapt according to the shape of hand in depth image.       
\section*{Acknowledgements}
This work was partially funded by the European project Eyes of Things (EoT) under contract number GA643924.  
{\small
\bibliographystyle{ieee}
\bibliography{egbib}
}

\end{document}